\documentclass[12pt]{article}

\usepackage{amssymb}
\usepackage[dvips]{epsfig}
\usepackage{graphicx}
\usepackage{amsfonts}
\usepackage[utf8]{inputenc}

\newcommand{\be}{\begin{equation}}
\newcommand{\ee}{\end{equation}}
\newcommand{\bn}{\begin{eqnarray}}
\newcommand{\en}{\end{eqnarray}}

\newcommand{\p}{\partial}

\newcommand{\nn}{\nonumber}
\newcommand{\no}{\noindent}

\newcommand{\s}{\,\,\,\,}
\def\bea{\begin{eqnarray}}
\def\eea{\end{eqnarray}}

\newcommand{\beq}{\begin{equation}}
\newcommand{\eeq}{\end{equation}}

\begin{document}

\title{\textbf{Duality of parity doublets of helicity $\pm 2$ in $D=2+1$}}
\author{D. Dalmazi  and Elias L. Mendon\c ca \\
\textit{{UNESP - Campus de Guaratinguet\'a - DFQ} }\\
\textit{{Av. Dr. Ariberto Pereira da Cunha, 333} }\\
\textit{{CEP 12516-410 - Guaratinguet\'a - SP - Brazil.} }\\
\textsf{E-mail: dalmazi@feg.unesp.br , elias.fis@gmail.com }}
\date{\today}
\maketitle

\begin{abstract}

In $D=2+1$ dimensions there are two dual descriptions of parity
singlets of helicity $\pm 1$, namely the self-dual model of
first-order (in derivatives) and the Maxwell-Chern-Simons theory
of second-order. Correspondingly, for helicity $\pm 2$ there are
four models $S_{SD\pm}^{(r)}$ describing parity singlets of
helicities $\pm 2$. They are of first-, second-,third- and
fourth-order ($r=1,2,3,4$) respectively. Here we show that the
generalized soldering of the opposite helicity models
$S_{SD+}^{(4)}$ and $S_{SD-}^{(4)}$ leads to the linearized form
of the new massive gravity suggested by Bergshoeff, Hohm and
Townsend (BHT) similarly to the soldering of $S_{SD+}^{(3)}$ and
$S_{SD-}^{(3)}$. We argue why in both cases we have the same
result. We also find out a triple master action which interpolates
between the three dual models: linearized BHT theory,
$S_{SD+}^{(3)} + S_{SD-}^{(3)}$ and $S_{SD+}^{(4)} +
S_{SD-}^{(4)}$. By comparing gauge invariant correlation functions
we deduce dual maps between those models. In particular, we learn
how to decompose the field of the linearized BHT theory in
helicity eigenstates of the dual models up to gauge
transformations.

\end{abstract}

\newpage

\section{Introduction}

In $D=2+1$ dimensions it is possible to have a local description
of a massive spin-1 particle by means of one vector field without
breaking gauge invariance. Such theory is called
Maxwell-Chern-Simons (MCS)  and it was introduced in \cite{djt}.
It is a second-order (in derivatives) model which describes a
parity singlet of helicity $+1$ or $-1$, according to the sign in
front of the Chern-Simons term. The MCS theory is invariant under
the usual $U(1)$ gauge transformations $\delta_{\xi}A_{\mu} =
\p_{\mu}\xi$. Another model, named self-dual (SD) model, was found
later in \cite{tpn}. It shares the particle content of the MCS
theory but it is of first-order and it has no local symmetries.
Part of the SD model, namely, the Chern-Simons term, is invariant
under $\delta_{\xi}A_{\mu}$. By means of a  Noether embedment of
this symmetry it is possible
 to obtain the MCS theory from the
SD model, see \cite{anacleto}.

A similar picture applies for spin-2 particles in $D=2+1$. A
third-order model, the so called topologically massive gravity,
was introduced in \cite{djt} to describe a gravitational theory
with a massive graviton of helicity $+2$ or $-2$, according to the
sign in front of the gravitational Chern-Simons term, without
breaking the general coordinate invariance of the Einstein-Hilbert
action. The linearized version of this model about a flat
background will be denoted here by $S_{SD\pm}^{(3)}$ respectively.
Later \cite{aragone}, a self-dual model of first-order
$S_{SD\pm}^{(1)}$, similar to its spin-1 counterpart \cite{tpn},
was introduced as well as a second-order model ($S_{SD\pm}^{(2)}$)
analogous to the MCS theory, see \cite{desermc}. Recently, a new
self-dual theory of fourth-order ($S_{SD\pm}^{(4)}$) has been
found \cite{sd4,andringa}. In \cite{sd4} we have shown that
starting with the lowest-order model $S_{SD\pm}^{(1)}$ there is a
natural sequence of Noether embedment of gauge symmetries  such
that $S_{SD\pm}^{(i)} \to S_{SD\pm}^{(i+1)}$ with $i=1,2,3$
culminates at $S_{SD\pm}^{(4)}$. The same reasoning applied on the
spin-1 case ($SD \to MCS$) terminates at the MCS theory. Both MCS
and $S_{SD\pm}^{(4)}$ consist of two terms invariant under the
same set of local symmetries. Thus, there is no symmetry left for
a further embedment. This indicates that those models might be the
highest-order models to describe particles of helicity $\pm 1$ and
$\pm 2$ respectively in terms of only one fundamental field.

On the other hand, in the spin-1 case, it is well known that the
Proca theory describes in $D=2+1$ a parity doublet of helicities
$+1$ and $-1$ which is the same particle content of two SD models
of opposite helicities. Since both models (pair of SD and Proca)
have no local symmetries one might wonder whether they could be
identified. In fact, it is easy to show \cite{solda2} that the
pair of SD models of opposite helicities corresponds to a first
order version of the Proca model after some trivial rotation.
However, regarding its dual theory, a pair of MCS models of
opposite helicities, it is not so easy to identify it with the
Proca theory due to the local $U(1)$ symmetry of the MCS theory.
An extra ``interference term'' between the opposite helicities is
needed to comply with the local symmetries. This extra term can be
produced by the soldering formalism \cite{stone} as shown in
\cite{bk,gsd}. The idea of fusing two fields representing
complementary aspects of some symmetry into one specific
combination of fields is the core of the soldering procedure, see
also \cite{amorim} and \cite{abw509}.

In the spin-2 case it is the Fierz-Pauli \cite{fierz} theory which
plays the role of the Proca theory. Once again it is possible to
show \cite{solda2} that the pair $S_{SD+}^{(1)} + S_{SD-}^{(1)}$
is equivalent, after a rotation, to a first-order version of the
Fierz-Pauli (FP) theory while the dual pair $S_{SD\pm}^{(2)}$ must
be soldered in order to furnish the FP theory.  Remarkably, the
soldering of a pair of third-order models $S_{SD\pm}^{(3)}$ does
not reproduce the FP theory and leads to a unitary \cite{oda}
fourth-order theory  describing a parity doublet of helicities
$+2$ and $-2$ just like the FP theory. This model corresponds
precisely to the linearized version of the recently proposed new
massive gravity theory \cite{bht}, henceforth LBHT theory. It is
therefore natural to try to solder also a pair of the top models
$S_{SD\pm}^{(4)}$. In the next section we carry this out and end
up again with the LBHT theory. This suggests the uniqueness of the
LBHT model as a unitary higher-derivative model describing a
parity doublet of helicities $\pm 2$ in $D=2+1$.

In previous examples of soldered second-order models for spin-1
\cite{bk,gsd} and spin-2 \cite{solda2} it turns out that the
theories before and after soldering can be shown to be equivalent
at quantum level. This has been shown in \cite{bkm,jhep2} and
\cite{dual}  by means of the master action technique \cite{dj}. In
the second part of this work (section 3) we define a triple master
action which interpolates between the linearized BHT theory,
$S_{SD+}^{(3)} + S_{SD-}^{(3)}$ and $S_{SD+}^{(4)} +
S_{SD-}^{(4)}$. Thus, proving the quantum equivalence of all three
models in agreement with the soldering predictions of
\cite{solda2} and section 2 of the present work. The introduction
of convenient source terms allow us to derive dual maps between
gauge  invariants of those theories.

\section{Soldering $S_{SD+}^{(4)}$ and $S_{SD-}^{(4)}$}

It is necessary to fix the notation before we go on. Throughout
this work indices are lowered and raised by the flat metric:
$\eta_{\alpha\beta}={\rm diag}\left(-,+,+\right)$. Inside
integrals we use a shorthand notation similar to differential
forms:

\be \int A \cdot d\, B \, \equiv  \int  d^3 x \, A^{\mu\alpha} \,
\epsilon_{\mu}\,^{\nu\lambda} \p_{\nu} B_{\lambda\alpha}
\label{csl1} \label{prod} \ee

\no Frequent use will be made of the rank two tensor
$\Omega_{\alpha}^{\s\gamma}(h)=\epsilon^{\gamma\mu\nu}\lbrack
\p_{\alpha}h_{\nu\mu}-
\p_{\mu}(h_{\nu\alpha}+h_{\alpha\nu})\rbrack$ and of the symmetric
and anti-symmetric operators
$\theta_{\mu\nu}=(\eta_{\mu\nu}-\p_{\mu}\p_{\nu}/\Box)$ and
$E^{\mu\nu}=\epsilon^{\mu\nu\alpha}\p_{\alpha}$ respectively.

Some of the actions here can be interpreted as quadratic
truncations (linearized versions) about a flat background. In
particular, with $g_{\mu\nu}=\eta_{\mu\nu} + h_{\mu\nu}$, the
linearized Einstein-Hilbert action (LEH), linearized gravitational
Chern-Simons (LGCS) term, and linearized $K$-term \cite{bht} can
be written respectively as:

\be  \int  d^3 x \, \sqrt{-g}R\Big\vert_{hh} = \int  d^3 x \,
h_{\mu\alpha}E^{\mu\lambda}
E^{\alpha\gamma}h_{\gamma\lambda}=-\frac 12 \int  d^3 x \, h \cdot
d\Omega(h), \label{leh} \ee

\be \frac{1}{2}\int  d^3 x \, \left\lbrack
\epsilon^{\mu\nu\rho}\Gamma^{\epsilon}_{\mu\gamma}\left(
\p_{\nu}\Gamma^{\gamma}_{\epsilon\rho}+\frac{2}{3}\Gamma^{\gamma}_{\nu\delta}
\Gamma^{\delta}_{\rho\epsilon}\right)\right\rbrack_{hh}=- \int d^3
x \, h_{\nu\mu}\Box\theta^{\nu\epsilon}
E^{\mu\rho}h_{\epsilon\rho}=\frac{1}{4}\int  d^3 x \,
\Omega(h)\cdot d\Omega(h), \label{lgcs} \ee

\bea \int  d^3 x \, \left\lbrack \sqrt{-g}\left(
R_{\mu\nu}R^{\mu\nu}-\frac{3}{8}R^{2}\right)\right\rbrack_{hh}&=&\frac{1}{2}
\int d^3 x \,
h_{\sigma\mu}\Box^2(2\theta^{\sigma\gamma}\theta^{\mu\nu}-\theta^{\gamma\nu}\theta^{\mu\sigma})
h_{\nu\gamma}\nn\\ &=& - \frac{1}{8} \int  d^3 x \, \Omega (h)
\cdot d\Omega(\Omega(h)) \quad . \label{lk}\eea

\paragraph*{\s\s\s\,\,} Now we start with a couple of new self-dual models recently obtained
in \cite{sd4,andringa}. Each model $S_{SD\pm}^{(4)}$ below, though
of fourth-order in derivatives is unitary \cite{unitary,andringa},
and describes one massive mode of mass $m_{\pm}$ and  helicity
$\pm 2$ in $D=2+1$ dimensions respectively. In a convenient
notation for the soldering approach we write:

\be S_{SD+}^{(4)}(A)= \int\,d^3x\,\left\lbrack
\frac{1}{4}A_{\rho\sigma}\Box^2(2\theta^{\rho\nu}\theta^{\sigma\mu}-\theta^{\rho\sigma}\theta^{\mu\nu})
A_{\mu\nu}+\frac{m_+}{2}A_{\lambda\mu}\Box\theta^{\lambda\alpha}E^{\mu\delta}A_{\alpha\sigma}\right\rbrack,\label{W1}\ee

\be S_{SD-}^{(4)} (B)= \int\,d^3x\,\left\lbrack
\frac{1}{4}B_{\rho\sigma}\Box^2(2\theta^{\rho\nu}\theta^{\sigma\mu}-\theta^{\rho\sigma}\theta^{\mu\nu})
B_{\mu\nu}-\frac{m_-}{2}B_{\lambda\mu}\Box\theta^{\lambda\alpha}E^{\mu\delta}B_{\alpha\sigma}\right\rbrack
.\label{W2}\ee

\no The tensor fields are symmetric
$A_{\alpha\beta}=A_{\beta\alpha} \, , \,
B_{\alpha\beta}=B_{\beta\alpha}$. The first term in both actions
above corresponds exactly to (\ref{lk}), and the second one is
proportional to the quadratic truncation of the gravitational
Chern-Simons term (\ref{lgcs}). As suggested in \cite{djt}, the
full nonlinear version of (\ref{lgcs}) together with the
Einstein-Hilbert action build up the so called topologically
massive gravity (TMG). Since the Einstein-Hilbert action is
substituted by the fourth-order $K$-term in $S_{SD\pm}^{(4)}$, we
may call such models a linearized higher derivative TMG.

Now let us recall the basic idea of the soldering procedure. The
actions (\ref{W1}) and (\ref{W2}) are invariant under independent
global shifts $\delta A_{\mu\nu} = \omega_{\mu\nu} \, ; \, \delta
B_{\mu\nu} = \tilde{\omega}_{\mu\nu}$. In the soldering procedure
\cite{bk,gsd,solda2} one lifts the global shift symmetry to a
local one and ties the fields $A_{\mu\nu}$ and $B_{\mu\nu}$
together by imposing that their local symmetry transformations are
proportional to each other:

\be \delta A_{\mu\nu}=\omega_{\mu\nu} \quad;\quad \delta
B_{\mu\nu}=\alpha\, \omega_{\mu\nu},\label{sym}\ee

\no where $\alpha$ is so far an arbitrary constant. From
(\ref{W1}),(\ref{W2}) and (\ref{sym})  we can write down:

\be \delta (S_{SD+}^{(4)}(A)+S_{SD-}^{(4)}(B))= \int \, d^3x\,
J^{\sigma}_{\alpha}\Box\theta^{\rho\alpha}\omega_{\rho\sigma}
\quad , \label{W1+W2}\ee

\no with the Noether-like current $J^{\sigma}_{\alpha}$ given by:

\be
J^{\sigma}_{\alpha}=\frac{\Box}{2}\left(2\theta^{\sigma\mu}C_{\mu\alpha}
- \eta_{\alpha}^{\,\sigma}\theta^{\mu\nu}C_{\mu\nu}\right)
+E^{\sigma\delta}D_{\alpha\delta} \quad . \label{J}\ee

\no Where we have used the following field combinations:

\be C_{\mu\nu} = A_{\mu\nu}+\alpha B_{\mu\nu}\quad;\quad
D_{\alpha\delta}=m_+A_{\alpha\delta}-\alpha m_-B_{\alpha\delta}.
\label{cd}\ee

\no At this point we may try to cancel the variation (\ref{W1+W2})
by the introduction of an auxiliary field $H^{\alpha}_{\sigma}$
with a specific variation $\delta
H^{\alpha}_{\sigma}=-\Box\theta^{\rho\alpha}\omega_{\rho\sigma}$
such that :

\be \delta\left(S_{SD+}^{(4)}(A)+S_{SD-}^{(4)}(B)+\int\,d^3x\,
J^{\sigma}_{\alpha} H^{\alpha}_{\sigma}\right)=\int\, d^3x\,
\delta J^{\sigma}_{\alpha}  H^{\alpha}_{\sigma}\label{var}\ee

\no Since

\be  \delta C_{\mu\nu} = \left( 1 + \alpha^2 \right)
\omega_{\mu\nu} \quad ; \quad \delta D_{\mu\nu} = \left( m_+ -
\alpha^2 m_-\right) \omega_{\mu\nu} \quad , \label{deltacd} \ee

\no we have

\bea \delta J^{\sigma}_{\alpha} &=& \frac{(1+\alpha^2)}2
\left(2\Box\theta^{\sigma\mu}\omega_{\mu\alpha} -
\eta_{\alpha}^{\,\sigma}\Box\theta^{\mu\nu}\omega_{\mu\nu}\right)
+ \left( m_+ - \alpha^2 m_-\right)
E^{\sigma\delta}\omega_{\alpha\delta}\nn \\
&=& -\frac{(1+\alpha^2)}2 \left(2\delta H^{\sigma}_{\alpha} -
\eta_{\alpha}^{\,\sigma}\delta H^{\mu}_{\mu}\right) + \left( m_+ -
\alpha^2 m_-\right) E^{\sigma\delta}\omega_{\alpha\delta}
\label{deltaj} \eea

\no In order to write the Lagrangian density on the right handed
side of (\ref{var}) as a local function of the auxiliary field
$H_{\sigma}^{\alpha}$ and its variation $\delta
H_{\sigma}^{\alpha}$ we are forced to choose

\be \alpha= \pm\sqrt{\frac{m_+}{m_-}} \label{alpha}\ee

\no which leads to the soldering action $S_S$ invariant under the
local transformations (\ref{sym}),

\be S_S=
S_{SD+}^{(4)}(A)+S_{SD-}^{(4)}(B)+\int\,d^3x\,\left\lbrack
H^{\alpha}_{\sigma}J^{\sigma}_{\alpha}+\frac{(1+\alpha^2)}{4}(2
H^{\alpha}_{\sigma}H^{\sigma}_{\alpha}- H^2)
\right\rbrack,\label{H}\ee

\no where $H=H^{\alpha}_{\alpha}$. Solving the algebraic equations
of motion of $H^{\beta}_{\nu}$ we can invert them in terms of
$J^{\sigma}_{\nu}$ and rewrite the expression (\ref{H}) as:

\be
S_S=S_{SD+}^{(4)}(A)+S_{SD-}^{(4)}(B)-\frac{1}{2(1+\alpha^2)}\int\,d^3x\,(J^{\sigma}_{\alpha}J^{\alpha}_{\sigma}-J^2)\label{j2}\ee

\no where $J=J^{\mu}_{\mu}$. The quadratic term in the Noether
current is interpreted \cite{bk,gsd} as an interference term
between the opposite helicity modes necessary to patch together
the actions $S_{SD+}^{(4)}$ and $S_{SD-}^{(4)}$ into a local
theory invariant under (\ref{sym}). Replacing $J^{\nu}_{\sigma}$
from (\ref{J}) in (\ref{j2}) we find:

\bea
S_S=S_{SD+}^{(4)}(A)+S_{SD-}^{(4)}(B)&-&\frac{1}{(1+\alpha^2)}\int\,d^3x\,\left\lbrack
\frac{1}{4}C_{\mu\nu}\Box^2(2\theta^{\alpha\mu}\theta^{\beta\nu}-\theta^{\mu\nu}\theta^{\alpha\beta})
C_{\alpha\beta}\right.\nn\\
&+&\left.
C_{\mu\nu}\Box\theta^{\sigma\mu}E^{\nu\gamma}D_{\sigma\gamma}-\frac{1}{2}D_{\alpha\nu}E^{\sigma\nu}E^{\alpha\gamma}
D_{\sigma\gamma}\right\rbrack.\eea

\no After some algebra it is possible to rewrite the soldered
Lagrangian density entirely in terms of the soldered field
$h_{\mu\nu}=\left(\alpha A_{\mu\nu}-B_{\mu\nu}\right)/\sqrt{m_+
m_-}$ which is invariant under the local shifts (\ref{sym}) with
$\alpha$ being any of the two possibilities given in
(\ref{alpha}), namely:

\bea{\cal{L}}_{S}&=&\frac{1}{(1+
\alpha^2)}\left\lbrack\frac{1}{4m_+m_-}h_{\mu\nu}\Box^2(2\theta^{\alpha\mu}\theta^{\nu\beta}-\theta^{\mu\nu}\theta^{\alpha\beta})
h_{\alpha\beta} -\frac{m_{+} -
m_{-}}{2m_+m_-}h_{\mu\nu}\Box\theta^{\sigma\mu}E^{\nu\gamma}h_{\sigma\gamma}\right.\nn\\
&-&\left.
\frac{1}{2}h_{\mu\nu}E^{\mu\alpha}E^{\nu\gamma}h_{\alpha\gamma}\Large\right\rbrack\label{hh}\eea

\no  By using $\alpha=\pm \sqrt{m_+/m_-}$ we can check that each
of the terms in (\ref{hh}) is invariant under the discrete
symmetry $(m_+,m_-)\rightarrow(-m_-,-m_+)$, which interchanges
$S_{SD+ }^{(4)}\leftrightharpoons S_{SD-}^{(4)}$. More
importantly, up to an overall constant, the Lagrangian ${\cal
L}_S$ corresponds precisely to the quadratic truncation of the
generalized ($m_+ \ne m_-$) new massive gravity theory of
\cite{bht}:

\be 2(1+\alpha^2){\cal{L}}_{S}=\left\lbrack
\sqrt{-g}R-\frac{m_+-m_-}{2m_+m_-}
\epsilon^{\mu\nu\rho}\Gamma^{\epsilon}_{\mu\gamma}\left(
\p_{\nu}\Gamma^{\gamma}_{\epsilon\rho}+\frac{2}{3}\Gamma^{\gamma}_{\nu\delta}
\Gamma^{\delta}_{\rho\epsilon}\right) -\frac{\sqrt{-g}}{m_+ m_
-}\left(R_{\mu\nu}R^{\mu\nu}-\frac{3}{8}R^2\right)\right\rbrack_{hh},\label{RR}\ee

\no This is a bit surprising, because we have found the same
soldered theory ${\cal L}_S$ in \cite{solda2} where we have
started with two third order self-dual models $S_{SD+}^{(3)}$ and
$S_{SD-}^{(3)}$ . This seems to indicate that the LBHT theory
might be the highest-order self-consistent (unitary) theory
describing a parity doublet of helicity $\pm 2$.

In order to get some clue on why the soldering of $S_{SD+}^{(4)}$
and $S_{SD-}^{(4)}$ leads to the same theory obtained from
$S_{SD+}^{(3)}$ and  $S_{SD-}^{(3)}$ we give below a rough
argument dropping the fields indices. The key point is some
freedom in defining the Noether current due to an integration by
parts. In both cases we can write:

\be \delta (S_{SD+}^{(r)}(A) + S_{SD+}^{(r)}(B)) = \int \, d^3x\,
J^{(r)} \p^p \omega \quad . \label{varr} \ee

\no Where $r=3,4$. The symbol $\p^p$ stands for some differential
operator of order $p$ whose explicit form is not important and may
be different in each expression. So $p$ simply counts the order of
some differential operator. Since the $S_{SD\pm}^{(r)}$ model
contains a term of order $r$ plus another one of order $r-1$, the
freedom to integrate by parts in (\ref{varr}) allows us to choose
any integer value for $p$ such that $p=0,1,\cdots ,r-1 $ and
redefine the Noether current accordingly:

\bea J^{(3)} &=& \p^{3-p} D + \p^{2-p} C \quad , \label{j3}
\\ J^{(4)} &=& \p^{4-p} C + \p^{3-p} D \quad , \label{j4} \eea

\no Where $C=A + \alpha\, B$ and  $D= m_+ A - \alpha \, m_- B$,
see (\ref{cd}). The term with odd number of derivatives in
$S_{SD\pm}^{(r)}$ carries the sign of particle's helicity and
gives rise to the D-combination in (\ref{j3}) and (\ref{j4}). The
formula (\ref{varr}) suggests the auxiliary field variation
$\delta H = - \p^p \omega $ which leads to, see (\ref{var}),

\be \delta \left( S_{SD+}^{(r)}(A) + S_{SD+}^{(r)}(B) + \int \,
d^3x\, J^{(r)} H   \right) = \int \, d^3x\, \delta J^{(r)} H \quad
, \label{deltas1} \ee

\no However, using (\ref{deltacd}) in (\ref{j3}) and (\ref{j4}) we
have:

\bea \delta J^{(3)} &=& - (m_+ - \alpha^2 m_-)\p^{3-2p} H -
(1+\alpha^2) \p^{2-2p} H \quad , \label{dj3}
\\ \delta J^{(4)} &=&  -
(1+\alpha^2) \p^{4-2p} H - (m_+ - \alpha^2 m_-)\p^{3-2p} H  \quad
. \label{dj4} \eea

\no Therefore, see (\ref{deltas1}), in order to avoid any dynamics
for the auxiliary field $H$ we must choose $\alpha^2=m_+/m_-$ in
both cases $r=3,4$ and p=1 for $r=3$ while $p=2$ if $r=4$ as we
have done in \cite{solda2} and here respectively. In fact, the
above argument holds also for the generalized soldering of
$S_{SD+}^{(2)}$ and $S_{SD-}^{(2)}$ carried out in \cite{solda2}
(see also \cite{ilha})  and the generalized soldering of two
Maxwell-Chern-Simons (MCS) theories of opposite helicities $\pm 1$
with different masses \cite{gsd} (see also \cite{bk}), in such
examples $p=0$. Finally, since in both cases $r=3,4$ we have
$\delta J^{(r)} = - (1 +\alpha^2)\delta \, H$ and the Noether
currents will be the sum of a 1st-order and a 2nd-order term, it
is clear that the interference term obtained after the elimination
of the auxiliary field will be quadratic in the current and can
only contain terms of order 4,3 and 2 which lead dimensionally to
the generalized BHT theory ${\cal L}_S$.

\section{Master action and dual maps}

In the soldering procedure there is {\it a priori} no guarantee of
quantum equivalence between the initial pair of field theories
describing the opposite helicity states and the final soldered
field theory. In the spin-1 case where a couple of MCS theories of
opposite helicities is soldered into a Maxwell-Chern-Simons-Proca
(MCSP) theory, even if $m_+\ne m_-$, it is possible to prove
 at quantum level the equivalence of those
theories before and after soldering by means of a master action
\cite{jhep2,bkm}. Likewise, in the spin-2 case one can also solder
\cite{solda2} the opposite helicities second-order models
$S_{SD+}^{(2)}$ and $S_{SD-}^{(2)}$ into a kind of spin-2 MCSP
model where the role of the Maxwell-Proca terms is played by the
Fierz-Pauli theory. Once again, those theories (before and after
soldering) are known to be quantum equivalent
 \cite{dual}. On one hand, such results are not surprising since the particle content of both
theories before and after soldering is the same, however the local
symmetries are in general not the same and the existence of a
local dual map between gauge invariant objects is not trivial.
From the above discussion and from what we have learned in the
last section it is quite suggestive to think about a master action
which interpolates among a couple of $S_{SD\pm}^{(4)}$, a couple
of $S_{SD\pm}^{(3)}$ and the LBHT theory. For simplicity we assume
hereafter $m_+=m_-$ and suggest the following master action:

\bea S_M \lbrack h,H,A,B \rbrack &=& \frac{1}{2}\int\, h\cdot
d\Omega(h)-\frac{1}{8 m^2}\int\, \Omega(h)\cdot
d\Omega(\Omega(h))\nn \\ &+&
\frac{1}{2}\int\,\left(H+\frac{\Omega(h)}{2m}\right)\cdot d\Omega
\left(H+\frac{\Omega(h)}{2m}\right)\nn\\
&+&\frac{1}{4m}\int\,\Omega(a-A)\cdot
d\Omega(a-A)-\frac{1}{4m}\int \Omega(b-B) \cdot
d\Omega(b-B).\label{sd4-1}\eea

\no Where all fields above are second-rank symmetric tensors with
$a_{\alpha\beta} $ and $b_{\alpha\beta} $ linear combinations of
$h$ and $H$ (dropping the indices):

\be a=\frac{(h+H)}{\sqrt{2}}\quad;\quad
b=\frac{(h-H)}{\sqrt{2}}.\label{subs}\ee

The first two terms in (\ref{sd4-1}) correspond to the LBHT
theory. Next, there are three mixing terms. The first one is a
quadratic truncation of the Einstein-Hilbert term, see
(\ref{leh}), while the last two ones are quadratic truncations of
the gravitational Chern-Simons term, see (\ref{lgcs}). All mixing
terms have no particle content and that feature plays a
fundamental role in the interpolation between the different models
\cite{jhep1,jhep2}. In order to verify the equivalence between
correlation functions of gauge invariants we are going to add a
source term to $S_M$. At this point we can ask what is the proper
source term. The fourth-order self-dual model is invariant under
linearized general coordinate transformations $\delta_{\xi}
h_{\mu\nu}= \p_{\mu}\xi_{\nu} + \p_{\nu}\xi_{\mu}$ and a
linearized local Weyl symmetry $\delta_{\phi}
h_{\mu\nu}=\phi\eta_{\mu\nu}$. On the other hand, the quadratic
Einstein-Hilbert term present in the LBHT and in $S_{SD\pm}^{(3)}$
breaks the local Weyl symmetry. The basic idea is to use a source
term invariant under a set of symmetries common to all models to
be interpolated. The lowest-order source term invariant under
$\delta_{\xi} h_{\mu\nu}$ is given by:

\be \int \, d^3x\, j^{\mu\nu}F_{\mu\nu}(h) \equiv \int \, d^3x\,
j^{\mu\nu}
E_{\mu}^{\gamma}E_{\nu}^{\lambda}\,h_{\gamma\lambda}=-\frac 12
\int \, j\cdot d\Omega (h)\label{source1}\ee

\no So for simplicity we first define the generating functional
with only one type of source:

\be {\cal{W}}[j]=
\int\,{\cal{D}}h_{\mu\nu}\,{\cal{D}}H_{\mu\nu}\,{\cal{D}}A_{\mu\nu}\,{\cal{D}}B_{\mu\nu}\,{\rm{exp}}\,i\left\lbrack
S_{M}(h,H,A,B)-\frac{1}{2}\int\, j\cdot
d\Omega(h)\right\rbrack,\label{eq2}\label{master2}\ee

It is easy to see that if we do the trivial shifts, dropping the
indices,  $A\to A+a$, $B\to B+b$ and $H\to H - \Omega(h)/2m$  in
(\ref{master2}) the last three terms of $S_M$ decouple completely
into three terms without particle content. Integrating over
$A_{\mu\nu}$, $B_{\mu\nu}$ and $H_{\mu\nu}$ we obtain up to an
overall constant:

\be
{\cal{W}}[j]=\int\,{\cal{D}}h_{\mu\nu}\,{\rm{exp}}\,i\left\lbrack
S_{LBHT}(h)-\frac{1}{2}\int\,j\cdot
d\Omega(h)\right\rbrack,\label{master3}\ee

\no Therefore, the spectrum of $S_M$ coincides with the one of the
quadratic truncation of the BHT theory for equal masses, i.e., a
parity doublet of helicities $\pm 2$ and mass ``$m$''. In the next
two sub-sections we are going to derive the dual models to LBHT
from (\ref{master2}).

\subsection{Duality between $S_{SD+}^{(3)} + S_{SD+}^{(3)}$ and the linearized BHT theory}

For a demonstration of equivalence of LBHT with one  couple of
third order self-dual models $S_{SD\pm}^{(3)}$, we rewrite the
first three terms of $S_M$. The generating functional
(\ref{master2}) becomes:

\bea
{\cal{W}}[j]&=&\int\,{\cal{D}}h_{\mu\nu}\,{\cal{D}}H_{\mu\nu}\,{\cal{D}}A_{\mu\nu}\,{\cal{D}}
B_{\mu\nu}\,{\rm{exp}}\,i\int \left\lbrack\frac{1}{2}
\, h\cdot d\Omega(h)+\frac{1}{2}\,H\cdot d\Omega(H)+\frac{1}{2m}\Omega(h)\cdot d\Omega(H)\nn\right.\\
&+&\left.\frac{1}{4m}\Omega(a-A)\cdot
d\Omega(a-A)-\frac{1}{4m}\Omega(b-B) \cdot
d\Omega(b-B)-\frac{1}{2}j\cdot
d\Omega(h)\right\rbrack\nn\\\label{w31}\eea

\no After the shifts $A\to A+ a$ and $B\to B+ b$ we can integrate
over $A$ and $B$ and get rid of the two third-order Chern-Simons
mixing terms which play no role in this subsection. Then,
inverting (\ref{subs}) we can decouple the fields  in (\ref{w31}).
Thus, the generating functional, up to an overall constant, can be
rewritten as:

 \bea
{\cal{W}}[j]&=&\int\,{\cal{D}}a_{\mu\nu}{\cal{D}}b_{\mu\nu}\,{\rm{exp}}\,i\left\lbrack
S_{SD+}^{(3)}(a)+S_{SD-}^{(3)}(b) -\frac{1}{2^{3/2}}\int\, j\cdot
d\Omega(a+b) \right\rbrack.\label{master5}\eea

\no Where

\be S_{SD\pm}^{(3)}(a) = -\int d^3 x \left\lbrack
a_{\mu\alpha}E^{\mu\lambda} E^{\alpha\gamma}a_{\gamma\lambda} \pm
\frac{1}{m}a_{\alpha\mu} \, \Box
\theta^{\alpha\gamma}E^{\beta\mu}a_{\gamma\beta}\right\rbrack \, .
\label{sd3pm} \ee

\no The first term represents the quadratic truncation of the
Einstein-Hilbert action with a negative sign, while the second one
is a similar truncation of the gravitational Chern-Simons action,
see (\ref{leh}) and (\ref{lgcs}) respectively. Deriving
(\ref{master3}) and (\ref{master5}) with respect to the source
$j^{\mu\nu}$ we have the following relationship between the
correlation functions:

\be \langle F_{\mu_1\nu_1}\lbrack h (x_1) \rbrack  \cdots
F_{\mu_N\nu_N}\lbrack h (x_N)\rbrack  \rangle_{LBHT} = \langle
\frac{F_{\mu_1\nu_1} \lbrack (a+b) (x_1)\rbrack }{\sqrt{2}} \cdots
\frac{F_{\mu_N\nu_N} \lbrack (a+b) (x_N) \rbrack }{\sqrt{2}}
\rangle_{S_{SD+}^{(3)}(a)+ S_{SD-}^{(3)}(b)}, \label{cf1} \ee

Consequently, the relevant gauge invariant quantity  in the LBHT
theory $F_{\mu\nu}\lbrack h(x)\rbrack  $ is given in terms of a
(gauge invariant) specific combination  of the fields with well
defined helicity: $F_{\mu_1\nu_1} \lbrack (a+b) (x_N) \rbrack
/\sqrt{2}$. However, for a complete proof of equivalence between
the decoupled pair $S_{SD\pm}^{(3)}$ and the linearized BHT theory
we should be able to compute correlation functions of
$F_{\mu\nu}\lbrack a (x)\rbrack$ and $F_{\mu\nu}\lbrack b
(x)\rbrack $ separately in terms of correlators of gauge invariant
objects in the LBHT theory. With this purpose in mind we define a
new generating function by changing the source term in
(\ref{master2}), i.e.,

\bea {\cal{W}}[j_+,j_-]&=&\int\,{\cal{D}}h_{\mu\nu}\,{\cal{D}}
H_{\mu\nu}\,{\cal{D}}A_{\mu\nu}\,{\cal{D}}B_{\mu\nu}\,{\rm{exp}}\,i\left\lbrace S_M(h,H,A,B)\nn\right.\\
&-&\left.\frac{1}{2}\int\,\left\lbrack j_+\cdot
d\Omega(h)+j_-\cdot
d\Omega(H)\right\rbrack\right\rbrace.\label{master4}\eea

\no The next steps will be totally equivalent to those we have
done previously, except for the fact that the source terms are now
redefined. Therefore we are going to suppress some details. Using
the same sequence of shifts that we have done from (\ref{master2})
to (\ref{master3}) we can verify that (\ref{master4}) after some
rearrangement is rewritten as:

\bea {\cal{W}}[j_+,j_-] &=&
\int\,{\cal{D}}h_{\mu\nu}\,{\cal{D}}H_{\mu\nu}\,{\rm{exp}}\,i\left\lbrace
S_{LBHT}(h)
-\frac{1}{2}\int\, \left\lbrack j_+\cdot d\Omega(h)-\frac{j_-\cdot d\Omega(\Omega(h))}{2m}\right\rbrack \right. \nn\\
&+&\left.\frac{1}{2}\int\left(H-\frac{j_-}2\right)\cdot
d\Omega\left(H-\frac{j_-}2 \right) - \frac{1}{8}\int j_- \cdot
d\Omega (j_-) \right\rbrace ,\label{above}\eea

\no shifting $H\to H + j_-/2$ in (\ref{above}), we can decouple
$H_{\alpha\beta}$ from the sources $(j_-)_{\alpha\beta}$ and
obtain an Einstein-Hilbert term for the field $H_{\alpha\beta}$
which has no particle content. Integrating over such field we
have, up to an overall constant,

\be {\cal{W}}[j_+,j_-]=
\int\,{\cal{D}}h_{\mu\nu}\,{\rm{exp}}\,i\left\lbrace
S_{LBHT}(h)-\frac{1}{2}\int\, \left\lbrack j_+\cdot
d\Omega(h)-\frac{j_-\cdot\Omega(\Omega(h))}{2m}\right\rbrack
+{\cal{O}}(j^2)\right\rbrace.\label{lnmg2}\ee

\no On the other hand, similarly to what we have done from
(\ref{w31}) to (\ref{master5}) we can write the expression for the
generating functional ${\cal{W}}[j_+,j_-]$ in terms of the
$S_{SD\pm}^{(3)}$ models as:

\begin{small}
\be {\cal{W}}[j_+,j_-]=
\int\,{\cal{D}}a_{\mu\nu}\,{\cal{D}}b_{\mu\nu}\,{\rm{exp}}\,i\left\lbrace
S_{SD+}^{(3)}(a)+S_{SD-}^{(3)}(b)-\frac{1}{2^{3/2}}\int\,
\left\lbrack j_+\cdot d\Omega(a+b)+j_-\cdot
d\Omega(a-b)\right\rbrack \right\rbrace.\label{soma2}\ee
\end{small}

\no The source terms in (\ref{soma2}) suggests the redefinition:

 \be \tilde{j}_+=\frac{j_++j_-}{\sqrt{2}}\quad ;\quad
\tilde{j}_-=\frac{j_+-j_-}{\sqrt{2}}.\label{rotations}\ee

\no which gives us:

\be {\cal{W}}[j_+,j_-]=
\int\,{\cal{D}}a_{\mu\nu}\,{\cal{D}}b_{\mu\nu}\,{\rm{exp}}\,i\left\lbrace
S_{SD+}^{(3)}(a)+S_{SD-}^{(3)}(b)-\frac{1}{2}\int\, \left\lbrack
\tilde{j}_+\cdot d\Omega(a)+\tilde{j}_-\cdot
d\Omega(b)\right\rbrack \right\rbrace.\label{soma3}\ee

\no Back in (\ref{lnmg2}) we have:

\bea \tilde{\cal{W}}[j_+,j_-]&=& \int\,{\cal{D}}h_{\mu\nu}\,{\rm{exp}}\,i\Bigg\{ S_{LNMG}(h)\nn\\
&-&\frac{1}{2^{3/2}m^2}\int\, \left\lbrack \tilde{j}_+\cdot
d\Omega\left(h-\frac{\Omega(h)}{2m}\right)+\tilde{j}_-\cdot
d\Omega\left(h+\frac{\Omega(h)}{2m}\right)\right\rbrack
+{\cal{O}}(\tilde{j}^2)\Bigg\}.\label{lnmg4}\eea

%

\no Deriving (\ref{soma3}) and (\ref{lnmg4}) with respect to the
sources $\tilde{j}_+$ and $\tilde{j}_-$ it is possible to map
correlations functions of the gauge invariant objects
$F_{\mu\nu}\lbrack a  (x)\rbrack$ and $F_{\mu\nu}\lbrack b
(x)\rbrack $ separately in terms of gauge invariants from the LBHT
theory as follows

 \bea  2^{\frac N2} \langle
F_{\mu_1\nu_1}\lbrack a  (x_1)\rbrack &\cdots &
F_{\mu_N\nu_N}\lbrack a (x_N) \rbrack \rangle_{S_{SD+}^{(3)}(a)}
\nn\\ &=& \langle \left( F_{\mu_1\nu_1} - G_{\mu_1\nu_1}\right)
\lbrack h(x_1) \rbrack \cdots \left( F_{\mu_N\nu_N} -
G_{\mu_N\nu_N}\right) \lbrack h(x_N) \rbrack
 \rangle_{LBHT}+ C.T \nn\\ \label{cf2}\eea

\bea  2^{\frac N2} \langle F_{\mu_1\nu_1}\lbrack b  (x_1)\rbrack
&\cdots & F_{\mu_N\nu_N}\lbrack b (x_N) \rbrack
\rangle_{S_{SD-}^{(3)}(b)} \nn\\ &=& \langle \left( F_{\mu_1\nu_1}
+ G_{\mu_1\nu_1}\right) \lbrack h(x_1) \rbrack \cdots \left(
F_{\mu_N\nu_N} + G_{\mu_N\nu_N}\right) \lbrack h(x_N) \rbrack
 \rangle_{LBHT}+ C.T \nn\\ \label{cf3}\eea

\no Where $C.T$ means contact terms which are due to the quadratic
terms in the sources while

\be G_{\alpha\beta}\lbrack a (x) \rbrack = -\frac{\Box}{m} \left(
E^{\rho}_{\,\alpha}\theta^{\delta}_{\, \beta} +
E^{\rho}_{\,\beta}\theta^{\delta}_{\, \alpha} \right)
a_{\rho\delta}(x) \quad  \label{G} \ee

\no is invariant under not only  linearized general coordinate
transformations $\delta_{\xi}a_{\rho\delta} =
\p_{\rho}\xi_{\delta} + \p_{\delta}\xi_{\rho}$ but also under
linearized Weyl symmetry $\delta_{\phi} a_{\rho\delta} = \phi \,
\eta_{\rho\delta}$ (use
$E^{\rho}_{\,\alpha}\theta_{\rho\beta}=E_{\beta\alpha}$).

 From (\ref{cf2}) and (\ref{cf3}) the dual maps are

\be F_{\mu\nu}\lbrack a  (x)\rbrack \Big{\vert}_{S_{SD+}^{(3)}(a)}
\longleftrightarrow \frac 1{\sqrt{2}}\left( F_{\mu\nu} -
G_{\mu\nu}\right) \lbrack h(x) \rbrack\Big{\vert}_{LBHT} \quad ,
\label{map1} \ee

\be F_{\mu\nu}\lbrack b  (x)\rbrack \Big{\vert}_{S_{SD-}^{(3)}(b)}
\longleftrightarrow \frac 1{\sqrt{2}}\left( F_{\mu\nu} +
G_{\mu\nu}\right) \lbrack h(x) \rbrack\Big{\vert}_{LBHT} \quad
.\label{map2} \ee

\no They are clearly consistent with (\ref{cf1}) and the
decomposition of $F_{\mu\nu}\lbrack h (x)\rbrack $ into the
linear combination of gauge invariant helicity eigenstates
$F_{\mu\nu} \lbrack (a+b) (x)\rbrack /\sqrt{2}$. The reader might
ask what happens when we subtract (\ref{map1}) from (\ref{map2}).
In this case we have  $F_{\mu\nu}\lbrack (a-b) (x)\rbrack $
calculated in the $S_{SD+}^{(3)}(a) + S_{SD-}^{(3)}(b)$ theory in
terms of $G_{\mu\nu}\lbrack h (x)\rbrack $ calculated in the
linearized BHT theory. If we recall that $\Box
\theta^{\mu}_{\,\alpha} = - E^{\mu\nu}E_{\nu\alpha}$ it is clear
from (\ref{G}) that $G_{\alpha\beta}\lbrack h (x)\rbrack $ can be
written as a first order differential operator applied on
$F_{\mu\nu}\lbrack h (x)\rbrack $. Therefore correlation functions
of $F_{\mu\nu}\lbrack (a -b) (x)\rbrack $ are given in terms of
correlation functions of $F_{\mu\nu}\lbrack (a + b)(x)\rbrack $
both calculated in the $S_{SD+}^{(3)}(a) + S_{SD-}^{(3)}(b)$
theory, though $a$ and $b$ are independent helicity eigenstates.
There is in fact no contradiction since we have a nontrivial
first-order differential operator relating both correlation
functions. This is typical of self-dual theories and it happens
also when we have a pair of spin-1 MCS theories of opposite
helicities, see formulae (3.9) and (3.10) of \cite{jhep2} for a
simpler example.

In summary, we have a complete equivalence between $S_{SD+}^{(3)}
+ S_{SD-}^{(3)}$ and $S_{LBHT}$. In the next subsection we show
how the third-order linearized gravitational Chern-Simons mixing
terms in the master action $S_M$ allow us to interpolate also
between the fourth-order models $S_{SD+}^{(4)} + S_{SD-}^{(4)}$
and $S_{LBHT}$.

\subsection{Duality between $S_{SD+}^{(4)} + S_{SD+}^{(4)}$ and the linearized BHT theory}

From (\ref{subs}) and the intermediate expression (\ref{w31}) we
get

\begin{small}
\bea
{\cal{W}}[j]&=&\int\,{\cal{D}}a_{\mu\nu}\,{\cal{D}}b_{\mu\nu}\,{\cal{D}}A_{\mu\nu}\,{\cal{D}}B_{\mu\nu}
\,{\rm{exp}}\,i\Bigg\{ S_{SD+}^{(3)}(a)+S_{SD-}^{(3)}(b)\nn\\
&+&\frac{1}{4m}\int\,\Omega(a-A)\cdot d\Omega(a-A)-\frac{1}{4m}
\int\,\Omega(b-B)\cdot d\Omega(b-B)-\frac{1}{2^{3/2}}
\int\,j\cdot d\Omega(a+b)\Bigg\},\nn\\
\label{37}\eea
\end{small}


\no  The factors in front of the linearized gravitational
Chern-Simons mixing terms in $S_M$ have been fine-tuned to cancel
the third-order terms of $S_{SD+}^{(3)}(a)+S_{SD-}^{(3)}(b)$.
After those cancelations and some rearrangements we get:

\bea
{\cal{W}}[j]&=&\int\,{\cal{D}}a_{\mu\nu}\,{\cal{D}}b_{\mu\nu}\,{\cal{D}}A_{\mu\nu}
\,{\cal{D}}B_{\mu\nu}\,{\rm{exp}}\,i\Bigg\{ S_{SD+}^{(4)}(A)+S_{SD-}^{(4)}(B)\nn\\
&+&\frac{1}{2}\int\,\left(a - \frac{\Omega(A)}{2m}\right)\cdot
d\Omega\left(a -
\frac{\Omega(A)}{2m}\right)+\frac{1}{2}\int\,\left(b+\frac{\Omega(B)}{2m}\right)\cdot
d\Omega
\left(b+\frac{\Omega(B)}{2m}\right)\nn\\
&-&\frac{1}{2^{3/2}}\int\,j\cdot d\Omega (a + b)\Bigg\} .
\label{38} \eea


\no It is easy to see that if we make $a \to a + \Omega(A)/2m$ and
$b\to b-\Omega(B)/2m$ we have

\bea {\cal{W}}[j]&=&\int\,{\cal{D}}a_{\mu\nu}\,{\cal{D}}b_{\mu\nu}
\,{\cal{D}}A_{\mu\nu}\,{\cal{D}}B_{\mu\nu}\,{\rm{exp}}\,i\left\lbrack
S_{+}^{(4)}(A)
+S_{-}^{(4)}(B)-\frac{1}{2^{5/2}m}\int\,j\cdot d\Omega(\Omega(A-B))\right.\nn\\
&+&\left.\frac{1}{2}\int \left(a- \sqrt{2}\, j\right)\cdot d\Omega
\left(a- \sqrt{2}\, j\right) + \frac{1}{2}\int \left(b- \sqrt{2}\,
j\right)\cdot d\Omega
\left(b- \sqrt{2}\, j\right) + {\cal{O}}(j^2) \right\rbrack. \nn\\
\label{39}\eea

\no After trivial shifts and  integrating over $a_{\alpha\beta}$
and $b_{\alpha\beta}$ fields we deduce up to an overall constant:

\be
{\cal{W}}[j]=\int\,{\cal{D}}A_{\mu\nu}\,{\cal{D}}B_{\mu\nu}\,{\rm{exp}}\,i\left\lbrack
S_{SD+}^{(4)}(A)+S_{SD-}^{(4)}(B) - \frac{1}{2^{5/2}m}\int\,j\cdot
d\Omega(\Omega(A-B))+
{\cal{O}}(j^2)\right\rbrack\label{derive2}\ee

\no Deriving (\ref{master3}) and (\ref{derive2}) with respect to
the source $j$ we obtain the following relationship between
correlation functions:

\begin{small}
\bea 2^{N/2}\langle F_{\mu_1\nu_1}\lbrack h (x_1)\rbrack
&\cdots & F_{\mu_N\nu_N}\lbrack h (x_N)\rbrack \rangle_{LBHT} \nn\\
&=&  \langle G_{\mu_1\nu_1}\lbrack (A-B) (x_1)\rbrack \cdots
G_{\mu_N\nu_N}\lbrack (A-B) (x_N)\rbrack
\rangle_{S_{SD+}^{(4)}(A)+S_{SD-}^{(4)}(B)} + C.T. \nn\\
\label{41}\eea
\end{small}

\no Now we go in the reverse direction and find correlation
functions mapping gauge invariant objects of $S_{SD+}^{(4)}(A)$
and $S_{SD-}^{(4)}(A)$ separately in gauge invariants of LBHT.
 Exactly as in the previous subsection, we replace the source term
 $\int j \cdot d\Omega (h)$  in (\ref{master2}) by  $\int j_+ \cdot d\Omega (h) + \int j_- d\cdot \Omega (H)$
 which on one hand leads to (\ref{lnmg4}) and on the other hand,
 following our previous steps, amounts to replace  (\ref{derive2}) by the generating functional:

\bea
{\cal{W}}[j_+,j_-]&=&\int\,{\cal{D}}A_{\mu\nu}\,{\cal{D}}B_{\mu\nu}
\,{\rm{exp}}\,i\left\lbrack S_{SD+}^{(4)}(A)+S_{SD-}^{(4)}(B)\right.\nn\\
&-&\left.\frac{1}{4 m}\int\,\tilde{j}_+\cdot d\Omega(\Omega(A)) +
\frac{1}{4 m}\int\, \tilde{j}_-\cdot
d\Omega(\Omega(B))+{\cal{O}}(j^2)\right\rbrack.\label{final}\eea

\no Finally, deriving (\ref{lnmg4}) and (\ref{final}) with respect
to the sources $\tilde{j_+}$ and $\tilde{j_-}$ we find :

\bea   2^{\frac N2}\langle G_{\mu_1\nu_1}\lbrack A  (x_1)\rbrack &
\cdots & G_{\mu_N\nu_N}\lbrack A (x_N) \rbrack
\rangle_{S_{SD+}^{(4)}(A)} \nn \\ &=&  \langle \left(
F_{\mu_1\nu_1} + G_{\mu_1\nu_1}\right) \lbrack h(x_1) \rbrack
\cdots \left( F_{\mu_N\nu_N} + G_{\mu_N\nu_N}\right) \lbrack
h(x_N) \rbrack
 \rangle_{LBHT}+ C.T \nn \\ \label{cf4}\eea

\bea    (-2)^{\frac N2}\langle G_{\mu_1\nu_1}\lbrack B
(x_1)\rbrack &\cdots &G_{\mu_N\nu_N}\lbrack B (x_N) \rbrack
\rangle_{S_{SD-}^{(4)}(B)} \nn\\ &=&  \langle \left(
F_{\mu_1\nu_1} - G_{\mu_1\nu_1}\right) \lbrack h(x_1) \rbrack
\cdots \left( F_{\mu_N\nu_N} - G_{\mu_N\nu_N}\right) \lbrack
h(x_N) \rbrack
 \rangle_{LBHT}+ C.T \nn \\ \label{cf5}\eea

\no The correlation functions (\ref{cf4}) and (\ref{cf5}) lead to
the gauge invariant maps

\be G_{\mu\nu}\lbrack A  (x)\rbrack \Big{\vert}_{S_{SD+}^{(4)}(A)}
\longleftrightarrow  \frac{\left( F_{\mu\nu} +
G_{\mu\nu}\right)}{\sqrt{2}} \lbrack h(x)
\rbrack\Big{\vert}_{LBHT} \quad , \label{map3} \ee

\be G_{\mu\nu}\lbrack B  (x)\rbrack \Big{\vert}_{S_{SD-}^{(4)}(B)}
\longleftrightarrow - \frac{\left( F_{\mu\nu} -
G_{\mu\nu}\right)}{\sqrt{2}} \lbrack h(x)
\rbrack\Big{\vert}_{LBHT} \quad ,\label{map4} \ee

\no which are consistent with (\ref{41}). Analogously to the dual
maps of previous subsection, if instead of subtracting we add
(\ref{map3}) and (\ref{map4}) we get a relationship between
correlation functions of $ G_{\mu\nu}\lbrack (A +B) (x)\rbrack $
in terms of correlation functions of a first-order differential
operator acting on $ G_{\mu\nu}\lbrack (A - B) (x)\rbrack $ which
is again typical of self-dual models. This completes the proof of
quantum equivalence between $S_{SD+}^{(4)} + S_{SD-}^{(4)}$ and
the LBHT theory. In particular, we have learned how to decompose
the gauge invariant sector of the LBHT theory in terms of (gauge
invariant) helicity eigenstates of $S_{SD\pm }^{(4)}$, namely
$F_{\mu\nu} \lbrack h(x) \rbrack$ corresponds to
$G_{\mu\nu}\lbrack (A - B) (x)\rbrack \sqrt{2}$. We remark that
each $S_{SD\pm}^{(4)}$ theory is invariant under linearized
general coordinate and Weyl transformations, so it is not
surprising that we have the  tensor $G_{\mu\nu}$, see (\ref{G})
and the comments below that formula, on the left handed side of
(\ref{cf4}) and (\ref{cf5}).

\section{Conclusion}

Although previous soldering of second-order $S_{SD\pm}^{(2)}$ and
third-order $S_{SD\pm}^{(3)}$ spin-2 parity singlets have led us
to second-order (Fierz-Pauli theory) and fourth-order (linearized
BHT theory) parity doublets respectively, we have shown in section
2 that the soldering of fourth-order singlets $S_{SD\pm}^{(4)}$
has brought us back to the linearized BHT model. We have
technically explained why this must be so.  This is an indication
that the linearized BHT model \cite{bht} is the highest-order
self-consistent (unitary) model which describes a parity doublet
of helicities $+2$ and $-2$. The reader can check that according
to the argument given at the end os section 2 if we had a
higher-derivative model $S_{SD\pm}^{(r)}$ with $r > 4$ then, we
could have after soldering another higher-derivative ($r>4$)
description of parity doublets of spin-2 in $D=2+1$. However, the
symmetry arguments given in  \cite{sd4} indicate that
$S_{SD\pm}^{(4)}$ might be the top (highest-order) derivative
model for parity singlets of spin-2. If this is really the case
the linearized BHT model is in fact the highest-order description
of parity doublets.

On the other hand, from the point of view of the local symmetries
the soldering of  $S_{SD+}^{(3)} + S_{SD-}^{(3)}$ into the
linearized BHT theory is more surprising than the soldering of
$S_{SD+}^{(4)} + S_{SD-}^{(4)}$ into the same theory, since in the
first case the two theories (before and after soldering) are
invariant under the same set of local symmetries (linearized
general coordinate transformation) while in the second one the
models $S_{SD\pm}^{(4)}$ are also symmetric under linearized local
Weyl transformations which calls for an extra term in the
soldering to get rid of  the Weyl symmetry. In the first case it
should be possible to simply add $S_{SD+}^{(3)} + S_{SD-}^{(3)}$
in order to obtain the linearized BHT theory after eventually some
trivial manipulations without adding extra terms. This is the case
of the two first-order self-dual models of spin-1 and spin-2 which
are known  \cite{solda2} to lead to its second order counterparts
(Proca and Fierz-Pauli theories respectively in first order form)
after a simple addition followed by a trivial rotation. So far we
have not been able to do it in the case of the models
$S_{SD\pm}^{(3)}$.

In section 3 we have written down a triple master action which
interpolates between all three models, i.e., $S_{SD+}^{(3)} +
S_{SD-}^{(3)}$, linearized BHT and $S_{SD+}^{(4)} +
S_{SD-}^{(4)}$. By introducing adequate source terms we have
derived identities involving correlation functions in the
different models allowing us to deduce a precise dual map, see
(\ref{map1}),(\ref{map2}),(\ref{map3}) and (\ref{map4}), between
the relevant gauge invariants of the different dual theories. No
specific gauge condition has ever been used. In particular, we
have been able to decompose a gauge invariant of the linearized
BHT model in terms of helicity eigenstates of both
$S_{SD\pm}^{(3)}$ and $S_{SD\pm}^{(4)}$. Putting our master action
(\ref{sd4-1}) together with the one defined in \cite{bht} relating
the Fiez-Pauli theory to the linearized BHT model, as well as
using the decomposition of the Fierz-Pauli model in terms of a
couple of $S_{SD\pm}^{(1)}$ models as given in \cite{solda2} we
can build a unifying description of all known dual versions of
field theories describing parity doublets of helicities $+2$ and
$-2$ in $D=2+1$. As remarked in \cite{jhep1}, the key ingredient
in the master action approach is the use of mixing terms without
particle content.

\section{Acknowledgements}

D.D. is partially supported by CNPq and E.L.M. is supported by
CAPES.

\end{document}